\documentclass[lengthestimate,tighten,amssymb,prb,aps,twocolumn,floatfix,tightenlines,superscriptaddress]{revtex4}
\usepackage{graphicx,subfigure,color}
\begin{document}
\title{Conditional phase shift from a quantum dot in a pillar microcavity}

\author{A.B.~Young}\email{A.Young@bristol.ac.uk}
\affiliation{Merchant Venturers School of Engineering, Woodland Road
Bristol, BS8 1UB}

\author{R. Oulton}\affiliation{Merchant Venturers School of Engineering, Woodland Road
Bristol, BS8 1UB}\affiliation{H.H. Wills Physics Laboratory, Tyndall Avenue,  Bristol BS8 1TL, UK}

\author{C.Y. Hu}\affiliation{Merchant Venturers School of Engineering, Woodland Road Bristol, BS8 1UB}

\author{A.C.T. Thijssen}\affiliation{H.H. Wills Physics Laboratory, Tyndall Avenue,  Bristol BS8 1TL, UK}

\author{C. Schneider}\affiliation{Technische Physik, Physikalisches Institut and Wilhelm Conrad R\"ontgen-Center for Complex Material Systems, Universit\"at W\"urzburg, Am Hubland, 97474 W\"urzburg, Germany}

\author{S. Reitzenstein}\affiliation{Technische Physik, Physikalisches Institut and Wilhelm Conrad R\"ontgen-Center for Complex Material Systems, Universit\"at W\"urzburg, Am Hubland, 97474 W\"urzburg, Germany}

\author{M. Kamp}\affiliation{Technische Physik, Physikalisches Institut and Wilhelm Conrad R\"ontgen-Center for Complex Material Systems, Universit\"at W\"urzburg, Am Hubland, 97474 W\"urzburg, Germany}

\author{S. H\"ofling}\affiliation{Technische Physik, Physikalisches Institut and Wilhelm Conrad R\"ontgen-Center for Complex Material Systems, Universit\"at W\"urzburg, Am Hubland, 97474 W\"urzburg, Germany}

\author{L. Worschech}\affiliation{Technische Physik, Physikalisches Institut and Wilhelm Conrad R\"ontgen-Center for Complex Material Systems, Universit\"at W\"urzburg, Am Hubland, 97474 W\"urzburg, Germany}

\author{A. Forchel}\affiliation{Technische Physik, Physikalisches Institut and Wilhelm Conrad R\"ontgen-Center for Complex Material Systems, Universit\"at W\"urzburg, Am Hubland, 97474 W\"urzburg, Germany}
%

\author{J.G.~Rarity}\affiliation{Merchant Venturers School of Engineering, Woodland Road
Bristol, BS8 1UB}\

\date{\today}

\begin{abstract}
Large conditional phase shifts from coupled atom-cavity systems are a key requirement for building a spin photon interface. This in turn would allow the realisation of hybrid quantum information schemes using spin and photonic qubits. Here we perform high resolution reflection spectroscopy of a  quantum dot resonantly coupled to a pillar microcavity. We show both the change in reflectivity as the quantum dot is tuned through the cavity resonance, and measure the conditional phase shift induced by the quantum dot using an ultra stable interferometer. These techniques could be extended to the study of charged quantum dots, where it would be possible to realise a spin photon interface. 
\end{abstract}
\maketitle

\noindent Long term storage (and processing) of quantum superpositions requires the development of Ôstatic qubitsÕ with long decoherence times, such as electron spins on charged quantum dots (QD's)\cite{PhysRevA.68.012310}. Photons can then be used as "flying" qubits to transfer and encode information between storage and processing nodes. An essential component is then a high-fidelity spin-photon interface, which can transfer information between photons and spins deterministically. In this paper we demonstrate the first steps towards such a device using a quantum dot strongly coupled to a pillar microcavity. We measure a reflection amplitude and macroscopic phase shift conditional on an (initially uncharged) QD being in resonance with the cavity. The conditional phase shift implies that a singly charged dot in a similar cavity will cause large polarization rotations conditional on the spin state of the electron\cite{PhysRevB.78.085307} leading to non-demolition measurement of spin and eventually to a spin-photon entangling device\cite{PhysRevB.80.205326, PhysRevB.78.125318}, or spin-photon interface.

The single photon has long been recognised as the ideal transporter of quantum information encoded as polarization\cite{rar-nat-3-481} or time superpositions\cite{gi-rmp-74-145}. There are several candidates for the storing of quantum information including the ground state spins such as found in trapped atoms\cite{cqed,langer:060502}, ions\cite{ci-prl-95-4091}, colour centres in diamonds\cite{PhysRevLett.92.076401}, or electron spins on charged QD's\cite{PhysRevA.68.012310}. It is the latter that is of interest here because of the comparative ease of incorporating them into a solid state cavity. This would facilitate deterministic transfer of quantum information between photon and spins. Such an spin-photon interface is key to building quantum repeaters\cite{waks:153601}, universal gates\cite{PhysRevLett.92.127902}, and eventually large scale quantum computers\cite{PhysRevA.78.032318}.

\begin{figure}
\center
\includegraphics[scale=0.25]{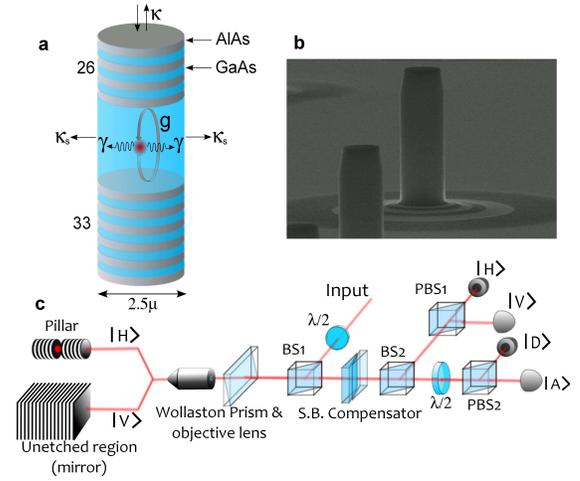}
\caption{{\bf a.} Showing a schematic diagram of a $2.5\mu$m dimeter pillar with 33/26 bottom/top mirror pairs, containing a single QD. {\bf b.} SEM image of a pillar microcavity defined by electron beam lithography and reactive ion etching. {\bf c.} Schematic diagram of the setup used to measure the phase shift caused by a single QD coupled to a pillar microcavity. BS1 and BS2 are both 50:50 non-polarising beamsplitters, the soleil babinet compensator allows us to initialise the setup with $sin\phi=0$. }\label{fig:schematic}
\end{figure}

The implementation of a practical conditional phase shift measurement and spin-photon interface in a QD-cavity system requires careful design considerations.  High quality-factor QD-cavities are now becoming routine, but if they are to be useful quantum information devices, one must collect the photons that eventually leak from the cavity with a reasonably high efficiency.  A high Q-factor photonic crystal slab cavity containing a QD has indeed been used to demonstrate a conditional phase shift\cite{IlyaFushman05092008} .  However, these types of cavities are designed either for low out-of-plane losses, or for high extraction efficiencies: it is difficult to achieve both high Q and high extraction efficiency in the same design.  Future designs may include an efficient coupled cavity-waveguide design to overcome this difficulty.  However, a spin-photon interface also requires a cavity that allows the readout of phase as an arbitrary superposition of linear polarization (i.e. one should be able to read out circular polarization without the cavity distorting the phase information).  No photonic crystal cavity design exists so far that would meet all of these requirements.  A cylindrical micropillar cavity, however, is much more promising.  The directionality of the emission may be fine-tuned by choosing the number of top distributed Bragg reflector (DBR) pairs, such that output efficiency is increased without compromising the Q-factor significantly, and its near-unity circular cross-section should lead to no birefringence in the mode of the cavity.  We demonstrate the feasibility of such a cavity for phase shift measurements, with an eye to an efficient spin-photon interface.

The planar microcavity structures are grown using molecular beam epitaxy on (100) orientation GaAs substrates. High reflectivity mirrors are fabricated by growing DBR's consisting of alternating layers of GaAs and AlAs of a quarter wave thickness, either side of a one wavelength thick GaAs cavity with a layer of AlGaInAs QDs formed in the cavity center. The sample used in this experiment consists of 33/26 bottom/top mirror pairs. Cylindrical micropillar devices are then fabricated using high resolution electron beam lithography and reactive ion etching\cite{reitzenstein:251109} (Fig.\ref{fig:schematic}(b)) with an eccentricity of $\epsilon<10^{-4}$. In this experiment the micropillar has a  nominal diameter of $2.5\mu$m.

The QD-cavity system can be parameterised by five constants (Fig.\ref{fig:schematic}(a)), these are: $\kappa$, the decay rate for intracavity photons via the top mirror (outcoupling), $\kappa_{s}$, the decay rate for intracavity photons via the side walls of the cylindrical micropillar, and any other losses such as transmission and absorption, $g$, the QD-cavity field coupling rate, and $\gamma$, the linewidth of the QD. We may now express the photon reflectivity when incident on the top of the pillar\cite{PhysRevB.78.085307}:

\begin{eqnarray}\label{eqn:ref}
&&r(\omega)=|r(\omega)|e^{i\phi}\\
&=&1-\frac{\kappa(i(\omega_{qd}-\omega)+\frac{\gamma}{2})}{(i(\omega_{qd}-\omega)+\frac{\gamma}{2})(i(\omega_{c}-\omega)+\frac{\kappa}{2}+\frac{\kappa_{s}}{2})+g^{2}}\nonumber
\end{eqnarray}

\noindent where $\omega_{qd}$ and $\omega_{c}$ are the frequencies of the QD and cavity, and $\omega$ is the frequency of incident photons. In this experiment we aim to find the reflectivity given by $|r(\omega)|^{2}$, and the associated phase shift $\phi$.

The setup used to perform the measurements can be seen in Fig.\ref{fig:schematic}(c). The waveplate is set on the input to give diagonally polarised ($+45^{\circ}$) light. This then passes through a Wollaston prism that separates the horizontal (H) and vertical (V) component at an angle of 0.5$^{\circ}$. We then use a lens with a numerical aperture (NA) of 0.5 to focus these to two spots ($\sim3\mu m$ diameter) on our sample; The H component is reflected from the cavity and the V component from some unetched material on the sample. The path difference between the two beams that form the two arms of the interferometer is $\sim$100$\mu$m, resulting in an ultra stable interferometric setup. The output signal is then split into two by a beamsplitter (BS2). On one arm the beam passes through a half waveplate set at $22.5^{\circ}$ which mixes the H and V components, then on to a polarising beam splitter (PBS1) to detect diagonal (D) and anti-diagonal (A). In the second arm we pass the signal through a polarising beamsplitter (PBS2) so that one channel is H and the other V. To obtain the phase we subtract the signal in detector "D" from that in detector "A" to give:

\begin{equation}\label{eq:interferometer}
D-A=r(\omega)\beta(\omega)sin\phi(\omega)
\end{equation}

\noindent where $r(\omega)$ is the amplitude of signal reflected from the pillar, and $\beta(\omega)$ is the amplitude of the signal reflected from the unetched region. Since the channels that measure "H" and "V" will give us $|r(\omega)|^{2}$ and $|\beta(\omega)|^{2}$ respectively, the signal can be normalised to extract the phase. This allows us to simultaneously measure both the intensity ($|r(\omega)|^{2}$), and phase ($\phi$) modulation of the QD coupled cavity system. 

To measure the amplitude and phase we perform high resolution reflection spectroscopy, by tuning a single frequency laser through the cavity resonance. The laser has a linewidth of $<1MHz$, and is attenuated to a power of $<100pW$ before the objective lens. This ensures we have $<0.1$ photons per cavity lifetime ($\sim25ps$). This enables very high resolution scans of the sample to be performed at the single photon level, where the QD transition is not saturated. We then subsequently change the temperature of the sample to tune the QD resonance through the cavity.

\begin{figure}
\center
\includegraphics[scale=0.5]{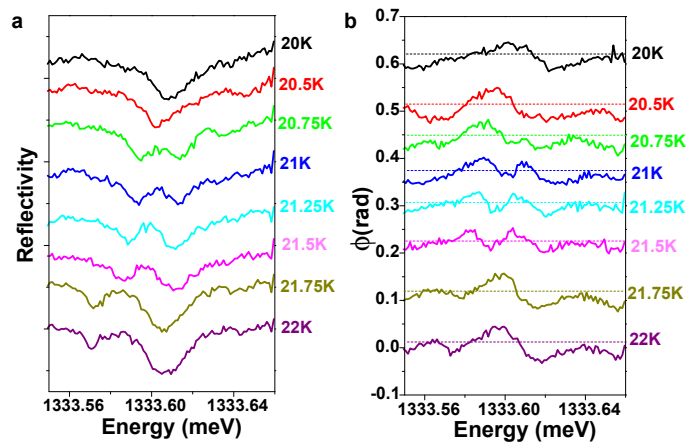}
\caption{Plots showing {\bf a}. the reflectivity of the QD micropillar, and {\bf b.} the phase as we vary the temperature of the sample. Since the dot energy changes quicker than the cavity energy we can tune the dot through the cavity resonance, from right to left as we increase temperature.}\label{fig:aandp}
\end{figure}

A QD-micropillar system was identified using non-resonant photo-luminesence (PL -not shown), in which a QD line (most likely a neutral exciton), temperature-shifted through the cavity without crossing. Fig.\ref{fig:aandp}(a) shows the change in reflectivity as a result of scanning the laser through the same cavity and QD resonance for different temperatures. At low temperature we have one dip in intensity caused by the cavity resonance with a Q-factor of $\sim51000$. As we increase the temperature to 20.5K the dot approaches from the higher energy side and can be seen to slightly modify the Lorentzian line shape. Further increasing the temperature we bring the dot closer to the cavity resonance, which forms a double peak structure due to the exiton-photon hybridisation. The dot then moves out of the other side of the cavity at higher temperatures, without ever crossing the cavity resonance. This anti crossing behaviour is a well known feature associated with strong coupling\cite{Reithmaier_nat}. At 22K the dot is still just visible in the reflection spectrum. 

In Fig.\ref{fig:aandp}(b) the corresponding measured phase shift is plotted as the dot is temperature tuned through the cavity. At low temperature there is only one phase feature associated with the cavity resonance, but as the QD is tuned onto resonance, a double phase feature associated with the Rabi split dressed states is observed. 

\begin{figure}\center
\includegraphics[scale=0.7]{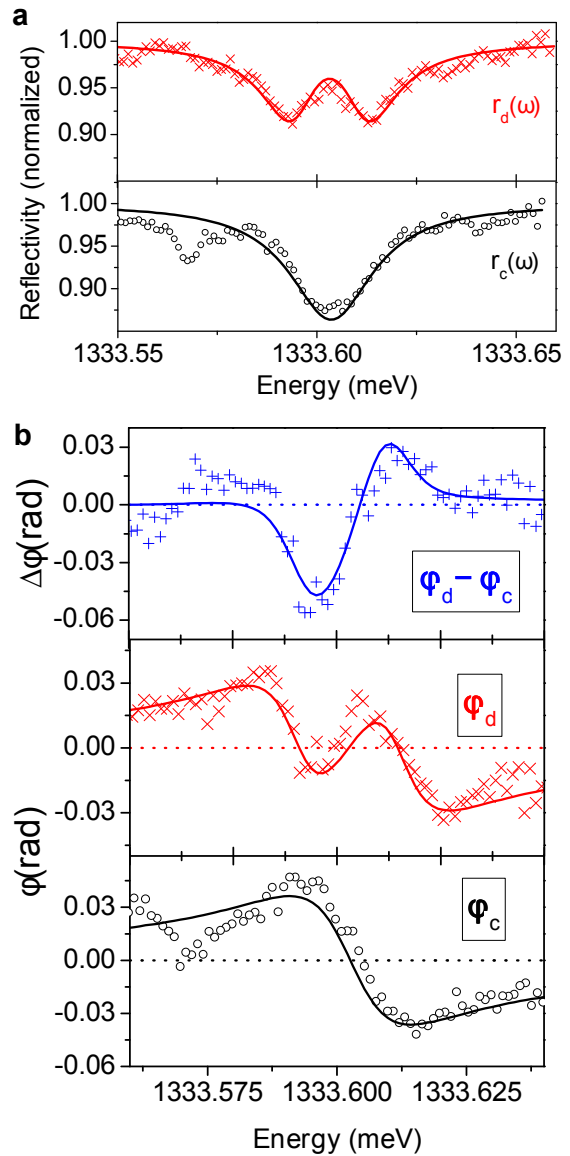}
\caption{Plots showing {\bf a.} The relfected amplitude for an empty cavity $r_{c} = |r(\omega)|_{c}^2$ at 22K, and a cavity with a resonantly coupled QD $r_{d}= |r(\omega)|^{2}_{d}$ at 21K. {\bf b.} Shows the corresponding phase shift for an empty cavity $\phi_{c}$, and a cavity with a resonantly coupled QD $\phi_{d}$, and the conditional phase shift $\phi_{d}-\phi_{c}$. the dashed lines represent the line of 0 phase shift for each plot.}\label{fig:aandpfit}

\end{figure}

Let us now examine the data in more detail. Fig.\ref{fig:aandpfit} shows data taken at 21K and 22K (QD on and off resonance) for (a) intensity and (b) phase. When the QD and the cavity are resonantly coupled ($\omega_{qd}=\omega_{c}$) at 21K, the double peak structure corresponds to the Rabi split dressed states. From this splitting it is possible to estimate $g\sim 11\mu eV$. Fitting the data using Equation \ref{eqn:ref} (Fig.\ref{fig:aandpfit}) we find that we obtain the values $g=9.4\mu$eV,  $\kappa_{s}= 24.7\mu$eV,  $\kappa=1.2\mu$eV and $\gamma\sim 5\mu$eV. This confirms that we are in the onset of strong coupling where $g>(\kappa+\kappa_{s}+\gamma)/4$. 

This fit also suggests that the photon loss rate ($\kappa_{s}$) is $\sim20$ times greater than that from the top DBR. However this ratio is determined by the depth of the features in the reflection spectra, which is also dependant on external mode coupling efficiency, not incorporated into Equation \ref{eqn:ref}. This coupling is dependant on the external optics: if we use a $\times100$ objective with a 0.7 NA ($<2\mu$m spot diameter) and scan the laser through the same bare cavity we find that the reflection feature decreases to $45\%$ of the maximum value (in supporting online material), compared to $15\%$ for the 0.5NA objective (Fig.\ref{fig:aandpfit}(a)). Assuming that at least a $3\times$ larger coupling efficiency is achievable, fitting this data implies that $\kappa_{s}\leq4\kappa$, reflecting previous estimates of the loss in these micropillar structures\cite{reitzenstein:251109}. By increasing the coupling efficiency we alter the ratio of $\kappa_{s}/\kappa$, however their sum remains constant and is the Q-factor of the cavity. As it is known that the modematching is still not optimum with the higher NA objective the ratio $\kappa_{s}/\kappa$ is likely more favourable than measured.

If we compare the results at 22K to those at 21K we observe clearly a conditional phase shift imposed on the reflected light from the resonantly coupled QD cavity ($\phi_{d}$) compared to the case of the empty cavity ($\phi_{c}$). Fig.\ref{fig:aandpfit}(b) contains a plot of this conditional phase shift ($\phi_{d}-\phi_{c}$). There is a maximum phase difference of 0.05rad ($2.85^{\circ}$) at 1333.596meV. The depth of the reflection feature in Fig.3(a) has a visibility of only $\sim0.15$, implying that the un-modematched reflected signal contributes a constant background of order $0.7$ to the reflected signal. Taking this into account we deduce a conditional phase shift of 0.12rad ($6.7^{\circ}$). Note that the phase shifts we infer are comparable to the non-linear phase shifts measured in previous work on photonic crystal QD-cavity systems ($12^{\circ}$)\cite{vuckovic:nat_450_857}, and atom cavity systems ($11^{\circ}$)\cite{tu-prl-75-4710}.  However, in our work the phase shift is linear and conditional on the QD being present in the cavity, not on the presence of a control photon. Additionally the value of the phase shift is measured directly: we do not filter the excitation photons that do not interact with the cavity.  This means that we are able to use the probe photons directly for use in quantum information applications.  If this were a charged dot, we would already have a giant Faraday rotation that is three orders of magnitude larger than has previously been measured for electron spins\cite{J.Berezovsky12222006}. Since measurements are performed at the single photon level it would be possible to perform a quantum non-demolition measurement of the spin.

Let us finally consider whether these results indicate that these state-of-the-art QD micropillar cavities are suitable for a spin-photon interface. One requirement for such an interface is a conditional phase shift $>\pi/2$ between an empty cavity and a resonantly coupled QD-cavity, this can be achieved if $\kappa > \kappa_{s}$.  We have a deduced phase shift of 0.12rad ($6.7^{\circ}$).  This could be increased by removing top mirror pairs from the DBR structure to increase $\kappa$. Calculations show if g and $\kappa_{s}$ remain constant and $\kappa$ is increased so that $\kappa>\kappa_{s}$, whilst maintaining $\kappa/4\sim g$ theoretically a conditional phase shift of $>\pi/2$ would be achieved. The disadvantage to working in this regime is that the calculated reflectivity is only $\sim20\%$. This would allow us to start performing the quantum operations proposed in\cite{PhysRevB.80.205326, PhysRevB.78.125318, PhysRevB.78.085307} albeit with a reduced efficiency due to intrinsic photon loss. Another requirement is an efficient input-output coupling, for deterministic transfer of information between photons and spins.  Here we have achieved of order $30\%$ in our phase measurement experiment, however with our $\times100$ objective we estimate this to be $>70\%$ (in supporting online material). In the future we would expect to achieve near $100\%$ efficiency as has previously been shown in similar pillar structures.\cite{PhysRevLett.102.097403} This would allow us to perform a significantly more accurate measurement of the side leakage and would also increase the observed phase shifts from the sample.

In conclusion we have shown that it is possible to measure the change in reflectivity from a micropillar caused by coupling a single QD, using high resolution reflection spectroscopy. We have also shown that we can deduce a conditional phase shift of $0.12$rad between an empty cavity and a cavity with a resonantly coupled QD, using a single photon level probe. With improved modematching and pillar design this conditional phase shift could be much larger. The intrinsic symmetry of the micropillar sample design means that the birefringence for linear polarizations can be minimal. This paves the way for future research using charged QDs, where, in the short term this technique would already allow us to monitor the spin dynamics of a single trapped electron spin strongly coupled to a cavity, with three orders of magnitude more accuracy than previously achieved.  This area is a crucial next-step to achieving the long-term goal of spin photon interfaces and deterministic quantum gates. 

\bibliographystyle{unsrtnat}

\begin{thebibliography}{22}
\providecommand{\natexlab}[1]{#1}
\providecommand{\url}[1]{\texttt{#1}}
\expandafter\ifx\csname urlstyle\endcsname\relax
  \providecommand{\doi}[1]{doi: #1}\else
  \providecommand{\doi}{doi: \begingroup \urlstyle{rm}\Url}\fi

\bibitem[Calarco et~al.(2003)Calarco, Datta, Fedichev, Pazy, and
  Zoller]{PhysRevA.68.012310}
T.~Calarco, A.~Datta, P.~Fedichev, E.~Pazy, and P.~Zoller.
\newblock Spin-based all-optical quantum computation with quantum dots:
  Understanding and suppressing decoherence.
\newblock \emph{Phys. Rev. A}, 68\penalty0 (1):\penalty0 012310, Jul 2003.

\bibitem[Hu et~al.(2008{\natexlab{a}})Hu, Young, O'Brien, Munro, and
  Rarity]{PhysRevB.78.085307}
C.~Y. Hu, A.~Young, J.~L. O'Brien, W.~J. Munro, and J.~G. Rarity.
\newblock Giant optical faraday rotation induced by a single-electron spin in a
  quantum dot: Applications to entangling remote spins via a single photon.
\newblock \emph{Phys. Rev. B}, 78\penalty0 (8):\penalty0 085307, Aug
  2008{\natexlab{a}}.



\bibitem[Hu et~al.(2009)Hu, Munro, O'Brien, and Rarity]{PhysRevB.80.205326}
C.~Y. Hu, W.~J. Munro, J.~L. O'Brien, and J.~G. Rarity.
\newblock Proposed entanglement beam splitter using a quantum-dot spin in a
  double-sided optical microcavity.
\newblock \emph{Phys. Rev. B}, 80\penalty0 (20):\penalty0 205326, Nov 2009.


\bibitem[Hu et~al.(2008{\natexlab{b}})Hu, Munro, and
  Rarity]{PhysRevB.78.125318}
C.~Y. Hu, W.~J. Munro, and J.~G. Rarity.
\newblock Deterministic photon entangler using a charged quantum dot inside a
  microcavity.
\newblock \emph{Phys. Rev. B}, 78\penalty0 (12):\penalty0 125318, Sep
  2008{\natexlab{b}}.


\bibitem[Ursin et~al.(2007)Ursin, Tiefenbacher, Schmitt-Manderbach, Weier,
  Scheidl, Lindenthal, Blauensteiner, Jennewein, Perdigues, Trojek, \"O~mer,
  F\"urst, Meyenburg, Rarity, Sodnik, Barbieri, Weinfurter, and
  Zeilinger]{rar-nat-3-481}
R.~Ursin, et al.
\newblock Entanglement-based quantum communication over 144 km.
\newblock \emph{Nature.}, 3:\penalty0 481 -- 486, June 2007.

\bibitem[Gisin et~al.(2002)Gisin, Ribordy, Tittel, and Zbinden]{gi-rmp-74-145}
Nicolas Gisin, Gr{\'{e}}goire Ribordy, Wolfgang Tittel, and Hugo Zbinden.
\newblock Quantum cryptography.
\newblock \emph{Rev. Mod. Phys.}, 74:\penalty0 145, 2002.

\bibitem[ed~P.~Berman(1994)]{cqed}
Kimble ed~P.~Berman.
\newblock \emph{Cavity Quantum Electrodynamics}.
\newblock Academic Press Inc, Boston, 1994.
\newblock p 203-266.

\bibitem[Langer et~al.(2005)Langer, Ozeri, Jost, Chiaverini, DeMarco, Ben-Kish,
  Blakestad, Britton, Hume, Itano, Leibfried, Reichle, Rosenband, Schaetz,
  Schmidt, and Wineland]{langer:060502}
C.~Langer, et al.
\newblock Long-lived qubit memory using atomic ions.
\newblock \emph{Physical Review Letters}, 95\penalty0 (6):\penalty0 060502,
  2005.


\bibitem[Cirac and Zoller(1995)]{ci-prl-95-4091}
J~I Cirac and P~Zoller.
\newblock Quantum computation with cold trapped ions.
\newblock \emph{Phys. Rev. Lett.}, 74:\penalty0 4091--4094, 1995.


\bibitem[Jelezko et~al.(2004)Jelezko, Gaebel, Popa, Gruber, and
  Wrachtrup]{PhysRevLett.92.076401}
F~Jelezko, T~Gaebel, I~Popa, A.~Gruber, and J~Wrachtrup.
\newblock Observation of coherent oscillations in a single electron spin.
\newblock \emph{Phys. Rev. Lett.}, 92\penalty0 (7):\penalty0 076401, Feb 2004.



\bibitem[Waks and Vuckovic(2006)]{waks:153601}
Edo Waks and Jelena Vuckovic.
\newblock Dipole induced transparency in drop-filter cavity-waveguide systems.
\newblock \emph{Physical Review Letters}, 96\penalty0 (15):\penalty0 153601,
  2006.


\bibitem[Duan and Kimble(2004)]{PhysRevLett.92.127902}
L.-M. Duan and H.~J. Kimble.
\newblock Scalable photonic quantum computation through cavity-assisted
  interactions.
\newblock \emph{Phys. Rev. Lett.}, 92\penalty0 (12):\penalty0 127902, Mar 2004.



\bibitem[Stephens et~al.(2008)Stephens, Evans, Devitt, Greentree, Fowler,
  Munro, O'Brien, Nemoto, and Hollenberg]{PhysRevA.78.032318}
Ashley~M. Stephens, et al.
\newblock Deterministic optical quantum computer using photonic modules.
\newblock \emph{Phys. Rev. A}, 78\penalty0 (3):\penalty0 032318, Sep 2008.








\bibitem[Fushman et~al.(2008)Fushman, Englund, Faraon, Stoltz, Petroff, and
  Vuckovic]{IlyaFushman05092008}
Ilya Fushman,et al.
\newblock {Controlled Phase Shifts with a Single Quantum Dot}.
\newblock \emph{Science}, 320\penalty0 (5877):\penalty0 769--772, 2008.
 

\bibitem[Englund et~al.(2007)Englund, Dirk, Andrei, Ilya, Nick, Pierre, and
  Jelena]{vuckovic:nat_450_857}
Englund, et al.
\newblock Controlling cavity reflectivity with a single quantum dot.
\newblock \emph{Nature}, 450:\penalty0 857--861, 2007.


\bibitem[Reitzenstein et~al.(2007)Reitzenstein, Hofmann, Gorbunov, Strauss,
  Kwon, Schneider, Loffler, Hofling, Kamp, and Forchel]{reitzenstein:251109}
S.~Reitzenstein, et al.
\newblock Alas/gaas micropillar cavities with quality factors exceeding
  150.000.
\newblock \emph{Applied Physics Letters}, 90\penalty0 (25):\penalty0 251109,
  2007.








%

\bibitem[Reithmaier et~al.(2004)Reithmaier, Sek, Loffler,  Hofmann, Kuhn, Reitzenstein,
 Keldysh, Kulakovskii, Reinecke and Forchel, A.]{Reithmaier_nat}
 Reithmaier et~al.
 \newblock{Strong coupling in a single quantum dot-semiconductor microcavity}
 \newblock \emph{Nature}, 432:\penalty0 197-200, 2004


\bibitem[Berezovsky et~al.(2006)Berezovsky, Mikkelsen, Gywat, Stoltz, Coldren,
  and Awschalom]{J.Berezovsky12222006}
J.~Berezovsky, et al.
\newblock {Nondestructive Optical Measurements of a Single Electron Spin in a
  Quantum Dot}.
\newblock \emph{Science}, 314\penalty0 (5807):\penalty0 1916--1920, 2006.

\bibitem[Turchette et~al.(1995)Turchette, Hood, Lange, Mabuchi, and
  Kimble]{tu-prl-75-4710}
Q~A Turchette, C~J Hood, W~Lange, H~Mabuchi, and H~J Kimble.
\newblock Measurement of conditional phase shifts for quantum logic.
\newblock \emph{Phys. Rev. Lett.}, 75:\penalty0 4710--4713, 1995.

\bibitem[Rakher et~al.(2009)Rakher, Stoltz, Coldren, Petroff, and
  Bouwmeester]{PhysRevLett.102.097403}
M.~T. Rakher, N.~G. Stoltz, L.~A. Coldren, P.~M. Petroff, and D.~Bouwmeester.
\newblock Externally mode-matched cavity quantum electrodynamics with
  charge-tunable quantum dots.
\newblock \emph{Phys. Rev. Lett.}, 102\penalty0 (9):\penalty0 097403, Mar 2009.


 \end{thebibliography}

\begin{small}

\noindent We acknowledge support from EPSRC EP/G004366/1, EU project 248095 Q-Essence and ERC grant 247462 QUOWSS. This work is carried out in the Bristol Centre for NanoScience and Quantum Information
\end{small}

\end{document}